\documentclass[usenatbib,twocolumn]{mnras} 
\usepackage[T1]{fontenc}
\usepackage{ae,aecompl}
\usepackage{graphicx}
\usepackage{hyperref}
\usepackage{amsmath}
\usepackage{amssymb}
\usepackage{mathtools}
\usepackage{bm}
\usepackage{cleveref}
\usepackage{widetext}
\usepackage{stfloats}

\usepackage{subcaption, caption}

\usepackage{float}
\usepackage{ dsfont }
\usepackage{ upgreek }
\usepackage{ulem,soul}

\usepackage{color,xcolor}

\title{Hybrid Stars Built with Density Dependent Models}
\author[A.~Issifu et al.]{
     A.~Issifu$^1$ 
    \thanks{E-mail: ai@academico.ufpb.br},     
    F. M. da Silva$^1$
    and
    D.~P.~Menezes$^1$ 
    \\ 
    \\
    $^1$Departamento de F\'{\i}sica - CFM - Universidade Federal de Santa Catarina  Florian\'opolis - SC - CP. 476 - CEP 88.040 - 900 - Brazil
    \\    }

\date{\today}

\begin{document}
    %
    %
    \pagerange{\pageref{firstpage}--\pageref{lastpage}} \pubyear{2023}
    \maketitle
    \label{firstpage}

\begin{abstract}
Using a density dependent quark model and a relativistic model within the mean-field approximation for hadrons with density dependent meson-baryon couplings, {we construct, for the first time, models that describe hybrid neutron stars} consisting of nucleons and exotic baryons (hyperons and $\Delta$-resonances). We do the study using a Maxwell construction. The quark-hadron phase transition in the stellar matter is determined through; the structure, composition, and properties of the hybrid neutron star matter. The macroscopic properties of the star are determined, and the results { for these particular models} are { found to be
compatible with recent} observational astrophysical data.

\end{abstract}

\begin{keywords}
Stars: Neutron, Stars: interiors
\end{keywords}


\section{Introduction}
Recent progress made in nuclear astrophysics due to the detection of gravitational waves from the merging of two neutron stars (NSs) in the event GW170817 \cite{LIGOScientific:2017vwq}, followed by the kilonova event observation in several wavelength bands of the electromagnetic spectrum \cite{Abbott_2017}, have given rise to the era of multi-messenger astronomy. These observations gave significant insight into the tidal deformability of merging NSs and provided new constraints on the equations of state (EoS) of these objects \cite{PhysRevD.81.123016}. Also, recent data from NICER \cite{Fonseca:2021wxt, Riley:2021pdl} gives a clear window for NS mass and radius. Besides, analysis of the  GW170817 merger signals also led to constraints on the radius of the NS \cite{PhysRevLett.120.172703, Bauswein:2017vtn} involved, giving hints about a possible phase transition in the core of the NS \cite{Bauswein_2019, PhysRevLett.125.141103}.

{
On the quantum chromodynamics (QCD) phase diagram, there is a line that separates hadronic matter from the quark-gluon plasma phase. A smooth crossover, confirmed by lattice QCD calculations, takes place
at high temperatures and low chemical potentials, giving rise to deconfinement. As the temperature decreases and the chemical potential increases, another phase transition, probably of the first order, appears from the hadronic to the quark phase. 
This is the likely scenario expected to occur in the inner core of the NSs \cite{PhysRevC.98.055805, Malfatti_2019}. }

The core of NSs is composed largely of strongly interacting protons and neutrons at low temperatures and high baryon density. However, moving deeper towards the inner core, heavy baryons such as the hyperons and the $\Delta$-resonances are expected to appear. Consequently, the baryons become tightly packed, such that, they may dissociate into a ``soup'' of deconfined quarks. Hence, { there is a possibility that there exist different phases of matter in the NS core: a hadronic phase at lower densities, a quark phase at higher densities, and perhaps even a mixture of hadrons and quarks \cite{LUKACS198727, PhysRevD.46.1274, PhysRevC.88.014906, Annala_2020}. }

The EoS for the NS matter in $\beta$-equilibrium is known to show different characteristics at two extreme limits. For densities $\sim\, 1.1\,n_0$, where the stellar matter exists with hadronic degrees of freedom, a chiral effective field theory can be used to calculate the EoS with good precision \cite{Gandolfi:2009fj, Tews:2012fj}. On the other hand, at higher densities, perturbative QCD techniques and high energy particle phenomenology, developed with quark-gluon degrees of freedom \cite{Kurkela:2009gj, Gorda:2018gpy}, give better results for the quark-matter EoS to an accuracy of about $n\,\gtrsim\,40n_0\,\equiv\,n_\text{pQCD}$. 

Observation gives indirect information about the matter content in the core of the NSs. Visualizing the matter content in NSs requires modeling strongly interacting matter from the crust to the highest density expected inside the star. There have not been an accurate prediction of matter phases in the core of the NS yet, due to the lack of first-principle predictions beyond the nuclear saturation density $n_0\,\approx\,0.152\,\text{fm}^{-3}$. However, gravitational wave data is unlikely to bring closure to this question shortly \cite{Bauswein:2018bma}. That notwithstanding, observational data has led to several strong constraints making model-independent approaches feasible. 

In this paper, the hadron matter is assumed to be composed of nucleons, hyperons, and delta isobars. The field equations for the hadrons are solved by adjusting them with the enhanced parameterization to the relativistic mean field (RMF) approximation method, known in the literature as DDME2 \cite{PhysRevC.71.024312}, and density dependent coupling constants determined from SU(3) symmetry arguments \cite{Lopes:2022vjx}. We also assume that the density at which quark deconfinement may take place is several times larger than the saturation baryon density of nuclear matter. We use the density dependent quark model (DDQM) to determine the EoS of the quark matter at the higher baryon density region where particles are expected to be in a deconfined state \cite{Backes:2020fyw, xia2014thermodynamic, wen2005thermodynamics}. {For the first time, two models with similar characteristics are used to describe the phase transition, namely,} the  DDQM together with DDME2. 
Thus, we can investigate the presence of a deconfined quark core in NS matter composed of an admixture of nucleons, hyperons, and $\Delta$-isobars. We ensure that the hybrid EoS developed is within the $2\,{\rm M}_\odot$ mass constraint \cite{PhysRevLett.120.172703}.


The paper is organized such that in Sec.~\ref{s1} we discuss the formalism of the EoSs used to construct the two-phase hybrid star EoS. The section is divided into two subsections; in Sec.~\ref{hadron} we discuss the hadronic EoS formulated from the RMF approach with density dependent baryon-meson couplings and in Sec.~\ref{quark} we discuss the DDQM used to calculate the quark matter EoS. In Sec.~\ref{s2} we discuss the deconfinement phase transition in a dense NS matter and how to visualize it from the EoSs of the hadronic and quark matter. We present a detailed analysis of the outcome of the investigation in Sec.~\ref{s3} and the final findings in Sec.~\ref{s4}.



\section{Equation of State}\label{s1}
In this section, we discuss separately the hadronic and quark matter models intended for use in constructing the two-phase model hybrid stars. We will elaborate on the RMF approximation with density dependent baryon-meson couplings and the DDQM in the following subsections. 

\subsection{RMF Approximation and Density Dependent Coupling}\label{hadron}
We study the hadronic region using quantum hydrodynamics (QHD-1) \cite{Walecka:1974qa, Serot:1984ey} (for review, see --- \cite{Menezes:2021jmw}), which describes particle interactions inside the nucleus in two forms: Long-distance attractive and short-distance repulsive interactions, used to describe confined and deconfined matter phases, respectively. The model is relativistic and commonly referred to as the RMF theory. It describes particle interactions as being mediated by mesons, as we will see shortly. The Lagrangian density of the model is
\begin{equation}
     \mathcal{L}_{\rm RMF}= \mathcal{L}_{H}+ \mathcal{L}_{\Delta}+ \mathcal{L}_{\rm mesons}+ \mathcal{L}_{\rm leptons},
\end{equation}
where $\mathcal{L}_{H}$, $\mathcal{L}_{\Delta}$, $\mathcal{L}_{\rm mesons}$ and $\mathcal{L}_{\rm leptons}$ are the Lagrangian densities for the baryon octet, baryon decuplet, mesons, and free leptons. The hadronic part is divided into the baryon octet and the decuplet. The Lagrangian of the baryon octet is of Dirac-type and takes the form
\begin{align}\label{lg}
 \mathcal{L}_{H}= {}& \sum_{b\in H}  \bar \psi_b \Big[  i \gamma^\mu\partial_\mu - \gamma^0  \big(g_{\omega b} \omega_0  +  g_{\phi b} \phi_0+ g_{\rho b} I_{3b} \rho_{03}  \big)\nonumber \\
 &- \Big( m_b- g_{\sigma b} \sigma_0 \Big)  \Big] \psi_b,
\end{align}
were $\sigma$ is a scalar meson, $\omega$ and $\phi$ are vector-isoscalar mesons and $\rho_{03}$ is an isovector-vector meson. The $\Delta$-isobars are represented by the Rarita-Schwinger-type Lagrangian 
\begin{align}\label{lg1}
        \mathcal{L}_{\Delta}={}& \sum_{d\in \Delta}\Bar{\psi}_{d\nu}\Big[\gamma^\mu i\partial_\mu- \gamma^0\left(g_{\omega d}\omega_0 + g_{\rho d} I_{3d} \rho_{03} \right) \nonumber\\&-\left(m_d-g_{\sigma d}\sigma_0 \right)\Big]\psi_{d\nu},
\end{align}
due to their additional vector-valued spinor component.
Even though it has been shown that the spin-3/2 and spin-1/2 models have the same equations of motion in the RMF approximation model \cite{dePaoli:2012eq}, it is still important to point out their distinctions. The mesonic part is represented by 

\begin{align}\label{lagrangian}
 \mathcal{L}_{\rm mesons}&= \dfrac{1}{2}\left(\partial_\mu\sigma\partial^\mu\sigma- m_\sigma^2 \sigma^2\right)  +\dfrac{1}{2}\left(\partial_\mu \omega\partial^\mu\omega- m_\omega^2 \omega^2 \right) \nonumber\\&+\frac{1}{2}\left(\partial_\mu\phi \partial^\mu\phi - m_\phi^2\phi^2\right) +\frac{1}{2}\left(\partial_\mu\vec{\rho}\partial^\mu\vec{\rho}-m_\rho^2 \vec{\rho}^2\right) .
\end{align}
Hereafter, we will consider only the mean field form for the analysis where $\sigma\rightarrow \langle\sigma\rangle \equiv \sigma_0$, $\omega\rightarrow \langle\omega\rangle 	\equiv \omega_0$ and $\Vec{\rho}\rightarrow \langle\rho\rangle 	\equiv \rho_{03}$. Finally, the leptons are represented by the free Dirac Lagrangian 
\begin{equation}\label{l1}
    \mathcal{L}_{\rm leptons} = \sum_L\Bar{\psi}_L\left(i\gamma^\mu\partial_\mu-m_L\right)\psi_L,
\end{equation}
where the summation runs over electrons (e) and muons ($\mu$) in the system, $L\in(e,\,\mu)$, and their antiparticles. The degeneracies of the leptons and the baryon octets are 2
while the degeneracy of the $\Delta$-isobars is 4.

We adjust the model with the enhanced density dependent parameterization, DDME2 \cite{PhysRevC.71.024312}, expressed as 
\begin{equation}
    g_{i b} (n_B) = g_{ib} (n_0)a_i  \frac{1+b_i (\eta + d_i)^2}{1 +c_i (\eta + d_i)^2},
\end{equation}
for $i=\sigma, \omega, \phi$ and 
\begin{equation}
    g_{\rho b} (n_B) = g_{ib} (n_0) \exp\left[ - a_\rho \big( \eta -1 \big) \right],
\end{equation}
for $i=\rho$, with $\eta =n_B/n_0$. The baryon-meson coupling adopted for this study was determined by \cite{Lopes:2022vjx} through SU(3) and SU(6) symmetry arguments, where the baryon-meson coupling for the $\Delta$-isobars were determined in a model-independent manner for the first time. The model parameters of the DDME2 were determined by fitting to experimental bulk nuclear matter data around its saturation density $n_0\,=\,0.152~ \text{fm}^{-3}$ and other properties such as the binding energy, compressibility modulus, symmetry energy, and slope \cite{PhysRevC.71.024312, Dutra:2014qga, Reed:2021nqk}. The values of the fit parameters are presented in Tab.~\ref{T1}. The ratio of the baryon to the nucleon coupling $\chi_{ib}=g_{ib}/g_{iN}$ with extension to the $\Delta$-isobars are shown in Tab.~\ref{T2}. 

The equations of motion of the meson fields are calculated using the Euler-Lagrange equation and solved together with the thermodynamic quantities of the baryons and the free non-interacting leptons, imposing $\beta$-equilibrium, charge neutrality, and baryon number conservation conditions. Under this description, the effective masses, and the effective chemical potentials of the particles are also density dependent. Detailed derivations of these quantities are contained in \cite{Roca-Maza:2011alv, Dutra:2014qga} for density dependent couplings.  The total energy density, $\varepsilon_B$, and pressure, $P_B$, are 
\cite{Issifu:2023qyi}:
\begin{flalign}\label{1a}
    \varepsilon_B&=  \varepsilon_b + \varepsilon_m + \varepsilon_d +\varepsilon_L ,\\
    P_B&=  P_b + P_m +P_d + P_L + P_r, \label{1b}
\end{flalign}
where the subscripts $b,\, d,\, L \, {\rm and }\, m$ represent baryon octet, $\Delta$-isobars, leptons and mesons, respectively. The pressure receives correction, $P_r$, due to thermodynamic consistency and energy-momentum conservation in the form
\begin{equation}
    P_r = n_B\Sigma^r,
\end{equation}
where $\Sigma^r$ is the rearrangement term \cite{Typel:1999yq, Fuchs:1995as}.  The effective masses $m^*_{b,d}$ and the effective chemical potentials $\mu^*$ of the baryons are 
\begin{equation}
    m_{b,d}^\ast  = m_{b,d} - g_{\sigma {b,d} } \sigma_0,
\end{equation}
and
\begin{align}
    \mu_{b,d}^\ast &= \mu_{b,d}- g_{\omega {b,d}} \omega_0 - g_{\rho {b,d}} I_{3{b,d}} \rho_{03} - g_{\phi {b}} \phi_0 - \Sigma^r,
\end{align}
respectively, with $I_{3{b,d}}$, the isospin projection. The baryon density is given by 
\begin{equation}
n_{b,d} = \gamma_{b,d} \int \frac{d^3 k}{(2\pi)^3},
\end{equation}
and the scalar density is 
\begin{equation}
    n^s_{b,d} =\gamma_{b,d} \int \frac{d^3 k}{(2\pi)^3} \frac{m^\ast_{b,d}}{E_{b,d}},
\end{equation}
where $\gamma_{b,d}$ is the particle degeneracy (2 for spin 1/2 particles and 4 for spin 3/2 particles) and $E_{b,d}= \sqrt{k^2 + m_{b,d}^{*2}}$ is the single particle energy.
 
\begin{table*}
\begin{center}
\begin{tabular}{ c c c c c c c }
\hline
 meson($i$) & $m_i(\text{MeV})$ & $a_i$ & $b_i$ & $c_i$ & $d_i$ & $g_{i N} (n_0)$\\
 \hline
 $\sigma$ & 550.1238 & 1.3881 & 1.0943 & 1.7057 & 0.4421 & 10.5396 \\  
 $\omega$ & 783 & 1.3892 & 0.9240 & 1.4620 & 0.4775 & 13.0189  \\
 $\rho$ & 763 & 0.5647 & --- & --- & --- & 7.3672 \\
 \hline
\end{tabular}
\caption {DDME2 parameters.}
\label{T1}
\end{center}
\end{table*}

\begin{table}
\begin{center}
\begin{tabular}{ c c c c c } 
\hline
 b & $\chi_{\omega b}$ & $\chi_{\sigma b}$ & $\chi_{\rho b}$ & $\chi_{\phi b}$  \\
 \hline
 $\Lambda$ & 0.714 & 0.650 & 0 & -0.808  \\  
$\Sigma^0$ & 1 & 0.735 & 0 & -0.404  \\
  $\Sigma^{-}$, $\Sigma^{+}$ & 1 & 0.735 & 0.5 & -0.404  \\
$\Xi^-$, $\Xi^0$  & 0.571 & 0.476 & 0 & -0.606 \\
  $\Delta^-$, $\Delta^0$, $\Delta^+$, $\Delta^{++}$   & 1.285 & 1.283 & 1 & 0  \\
  \hline
\end{tabular}
\caption {The ratio of the baryon coupling to the corresponding nucleon coupling for hyperons and $\Delta$s.}
\label{T2}
\end{center}
\end{table}
{The particle fraction of each constituent of the system is given by: }
\begin{equation}
    Y_i=\frac{n_i}{n_B} \label{yb}
\end{equation}
{where $i = {b,\, d}$ is related to the different particles}  {and the total baryon density
\begin{equation}
    n_B = \sum_{i}n_{i},
\end{equation}
is summed over all the baryons in the stellar matter.}

\subsection{Density Dependent Quark Model}\label{quark}
In this work, we assume that the deconfined matter phase in the inner core of the compact star is composed of electrons (e) and quarks (up ($u$), down ($d$), and strange ($s$)). In this phase, the matter will be described by the DDQM model expressed as
\begin{equation}
    m_i = m_{i0} + m_{I},
    \label{eq1}
\end{equation}
where $m_{i0} (i  = u,\, d,\, s)$ is the current quark mass and $m_{I}$ is the density dependent quantity which includes the interactions of the quarks. The motivation for the DDQM approach is to include the interactions between quarks in a simple way. It is important to consider how the interactions are included in (\ref{eq1}), details of introducing $m_{I}$ and some other parameterizations were discussed in \cite{fowler1981confinement,chakrabarty1989strange,peng1999mass}. Here, we consider the parameterization for the density dependent quark masses proposed in \cite{xia2014thermodynamic}, given by
\begin{equation}
    m_i = m_{i0} + \frac{D}{n_B^{1/3}} + C n_B^{1/3},
    \label{eq2}
\end{equation}
where $n_B$ is the baryon number density defined in (\ref{nB}), and $C$ and $D$ are the parameters of this model. The second term in (\ref{eq2}) is associated with linear confinement. Therefore, $D$ is a low-density parameter whose value is model dependent. The third term in (\ref{eq2}) is also associated with the leading-order perturbative interactions, which dominate at the higher-density regions; its value is model dependent. Some estimates for $C$ and $D$ were determined in \cite{Backes:2020fyw}, where they used stability as their benchmark for analysis. Also, in \cite{wen2005thermodynamics}, they found an estimate based on the relation between $C$ and $D$ and other physical quantities, such as the relation between $D$ and string
tension and the chiral restoration density. Besides in \cite{chen2021strangelets}, estimates were made based on stable radius. 
In this work, we use $C=0.965$ and $\sqrt{D}=121$ MeV which were {taken from \cite{Backes:2020fyw}, and are} within the unstable quark matter region in the model framework. Otherwise, the hybrid star is likely to transform into a strange quark star within a short time of its existence. {The other reason that informed this choice was that it maximizes the quark core.} The quark masses used are $5\,{\rm MeV}$, $10\,{\rm MeV}$ and $80\,{\rm MeV}$ for $u$, $d$ and $s$ quarks, respectively.

In addition, to calculate the EoS for the quark matter we impose a charge neutrality condition given by the following expression:
\begin{equation}
    \frac{2}{3} n_u - \frac{1}{3} n_d - \frac{1}{3} n_s - n_e = 0,
\end{equation}
where $n_i (i  ={\rm u, d, s, e})$ are the number densities of each particle. The quarks and electrons interact via weak interactions, which can produce neutrinos. However, as we are studying cold stars $(T = 0$ MeV $)$, the chemical potential of these neutrinos can be set to zero, and the chemical potential of the quarks and electrons obey the $\beta$-equilibrium condition
\begin{equation}
    \mu_u + \mu_e = \mu_d = \mu_s.
    \label{betaeq}
\end{equation}
Moreover, the relationship between the net baryon density and the quark number densities is given by
\begin{equation}\label{nB}
    n_B = \frac{1}{3} (n_u +n_d +n_s).
\end{equation}
{Here, the particle fraction of the quarks can be calculated using the expression above for $n_B$ and (\ref{yb}).}
Incorporating a density dependency can lead to thermodynamic inconsistencies, and one way to avoid this is by including an effective chemical potential $\mu^*_i$. Therefore, we can express the free-energy density $f$ of the free particle system with masses $m_i (n_B)$ and effective chemical potentials $\mu^*_i$
\begin{equation}
    f = \Omega_0 \left( \{ \mu^*_i \}, \{ m_i \} \right) + \sum_i \mu^*_i n_i,
\end{equation}
where $\Omega_0$ is the thermodynamic potential density of the free quarks with masses $m_i$ given by (\ref{eq1}), and effective chemical potential $\mu^*_i$. At $T=0$ MeV, $\Omega_0$ is given by the following expression
\begin{equation}
    \Omega_0 = - \sum_i \frac{g_i}{24 \pi^2} \left[ \mu^*_i \nu_i \left( \nu_i^2 -\frac{3}{2} m_i^2 \right) +\frac{3}{2} m_i^4 \ln{\frac{\mu^*_i +\nu_i}{m_i}}\right]
\end{equation}
where $g_i=6=(3$ colors $\times \; 2$ spins$)$ is the degeneracy factor, and $\nu_i$ are the Fermi momenta, which are now connected to the effective potentials by
\begin{equation}
    \nu_i = \sqrt{\mu^{*2}_i - m_i^2}.
\end{equation}
Thus, the particle number density $n_i$ is given by
\begin{equation}
    n_i = \frac{g_i}{6 \pi^2} (\mu^{*2}_i - m_i^2)^{3/2} = \frac{g_i \nu_i^3}{6 \pi^2}.
\end{equation}
The chemical potential $\mu_i$ and the effective chemical potential are related through the relation
\begin{equation}
    \mu_i = \mu_i^* - \mu_I,
\end{equation}
where $\mu_I$ is the density dependent quantity. Now, we can rewrite the $\beta$-equilibrium condition of Equation (\ref{betaeq}) as
\begin{equation}
    \mu_u^* + \mu_e = \mu_d^* = \mu_s^*.
\end{equation}
Lastly, we can express the energy density $\varepsilon_q$ and pressure $P_q$ of the particles as 
\begin{equation}
    \varepsilon_q = \Omega_0 - \sum_i \mu_i^* \frac{\partial \Omega_0}{\partial \mu_i^*},
\end{equation}
and
\begin{equation}
    P_q = -\Omega_0 + \sum_{i,j} \frac{\partial \Omega_0}{\partial m_j} n_i \frac{\partial m_j}{\partial n_i}.
\end{equation}


\section{Deconfinement Phase Transition and Hybrid EoS}\label{s2}

\begin{figure}
  \includegraphics[scale=0.5]{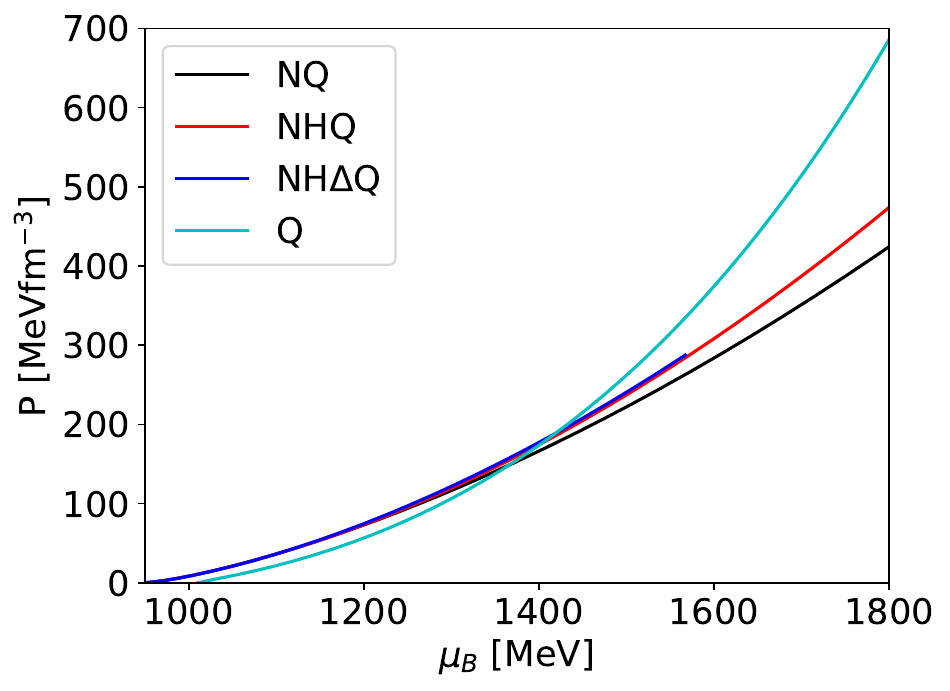}
\caption{Pressure as a function of baryon chemical potential.}    
\label{fig1}
\end{figure}

\begin{table}
\begin{center}

\begin{tabular}{l|l|l|l}
\hline \\
      Star Properties &  NQ & NHQ  & NH$\Delta$Q\\ \hline
$M_{\text{max}}[M_{\odot}]$ & \begin{tabular}[c]{@{}l@{}}2.22\end{tabular}  & \begin{tabular}[c]{@{}l@{}}2.19\end{tabular} & \begin{tabular}[c]{@{}l@{}}2.15\end{tabular} \\ \hline
$R[km]$ & \begin{tabular}[c]{@{}l@{}}13.03\end{tabular} & \begin{tabular}[c]{@{}l@{}}13.01\end{tabular} & \begin{tabular}[c]{@{}l@{}}12.55\end{tabular} \\ \hline
$\mu^H = \mu^Q[{\rm MeV}]$ & \begin{tabular}[c]{@{}l@{}}1364\end{tabular}  & \begin{tabular}[c]{@{}l@{}}1400\end{tabular} & \begin{tabular}[c]{@{}l@{}}1413\end{tabular}\\ \hline
 $P^H = P^Q [{\rm MeV}{\rm fm}^{-3}]$ & \begin{tabular}[c]{@{}l@{}}151.94\end{tabular}  & \begin{tabular}[c]{@{}l@{}}181.54\end{tabular} & \begin{tabular}[c]{@{}l@{}}185.49\end{tabular}\\
 \hline
$\varepsilon_0[{\rm MeV}{\rm fm}^{-3}]$ & \begin{tabular}[c]{@{}l@{}}838.64\end{tabular}  & \begin{tabular}[c]{@{}l@{}}826.80\end{tabular} & \begin{tabular}[c]{@{}l@{}}962.96\end{tabular}\\
 \hline
 $\varepsilon_H[{\rm MeV}{\rm fm}^{-3}]$ & \begin{tabular}[c]{@{}l@{}}562.38\end{tabular}  & \begin{tabular}[c]{@{}l@{}}670.51\end{tabular} & \begin{tabular}[c]{@{}l@{}}688.25\end{tabular}\\
 \hline
 $\varepsilon_Q[{\rm MeV}{\rm fm}^{-3}]$ & \begin{tabular}[c]{@{}l@{}}755.13\end{tabular}  & \begin{tabular}[c]{@{}l@{}}840.38\end{tabular} & \begin{tabular}[c]{@{}l@{}}898.34\end{tabular}\\
 \hline
\end{tabular}
\caption{Star properties}
\label{ma}
\end{center}
\end{table}

We develop the hybrid EoS using the two-phase model approach, where the hadronic and quark matter models are determined separately, and a hybrid EoS constructed through a phase transition. To determine the phase transition between hadronic and quark matter, we assume that the phase transition is first-order and Maxwell-like. 
In this scenario, the pressure in the {trasition} phase is constant. On the other hand, Gibbs-like phase transition is also well-known in the literature. 
However, it was determined that there are no significant differences between Maxwell and Gibbs constructions considering the microscopic properties of the hybrid stars \cite{Maruyama:2007ey, Paoli:2010kc}.  We adopted the Maxwell-like construction for this study.


Proceeding from the hadron and the quark models presented in Secs.~\ref{hadron} and \ref{quark}, we construct the hybrid EoSs for the stars by determining the EoS in the form of pressure as a function of chemical potential. We determine the crossing point between the hadronic and quark matter EoSs where the phase transition is energetically favorable. The crossing points can be seen in Fig.~\ref{fig1}, and the corresponding critical chemical potential and pressure are in Tab.~\ref{ma}. {Throughout the text, we have used the convention: the Q index represents the EoS for the DDQM model only and the NQ, NHQ, NH$\Delta$Q represent the three cases of hybrid EoSs that we have studied. NQ is the index for hybrid EoS where the pure hadronic phase is composed only of nucleons, NHQ means that the hadronic phase has nucleons plus hyperons and NH$\Delta$Q represents the case where the hadronic phase contains nucleons plus hyperons plus $\Delta$-isobars admixture.} The critical baryochemical potential, $\mu_c$, and the critical pressure $P_c$ at which the hadron and the quark matter are in mechanical and chemical equilibrium with each other is determined as 
\begin{equation}
    P_H(\mu) = P_Q(\mu) = P_c, \quad\text{and}\quad \mu_Q=\mu_H=\mu_c.
\end{equation}
This is shown in Fig.~\ref{fig1} where the curves for $P_H(\mu)$ and $P_Q(\mu)$ intersect. It is important to state that, at $\mu_c$, the quark phase becomes energetically favorable, and the hadron-quark phase transition occurs. The value of $\mu_c$ depends on the model employed, and its value indicates the point where hadronic and quark matter has the same chemical potential and pressure \cite{Lopes:2021jpm}. In this form, the lower baryon density region is composed of hadrons, and the higher-density region, where particles are in the deconfined state is described by the quark matter. In this construction, we ensure that the causality condition which forbids that the adiabatic speed of sound at zero frequency, $c_s\, =\, \sqrt{\partial P/\partial \varepsilon}$ does not exceed the constant speed of light.

{ It is important to note that the study of hybrid stars is certainly model dependent. Largely because their existence has neither been confirmed experimentally nor by observation. This topic has been widely discussed in the literature \cite{voskresensky2002charge,menezes2003warm,Maruyama:2007ey,Paoli:2010kc,Lopes:2020rqn,Lopes:2021yga,Lopes:2021jpm}. As an example, we can mention the study made in \cite{Lopes:2021jpm}, where different types of MIT-based models were analyzed, 
and it was shown clearly how these distinct models affect the results. In our case, different choices for the parameters C and D of the DDQM model, for example, can also change the results. In \cite{Backes:2020fyw} the authors showed mass-radius diagrams for different choices of C and D and the results significantly vary. Besides, different choices of hyperon or $\Delta$-baryon potentials would also change the results, as one can see in \cite{Marquez:2022gmu}. However, it is important to emphasize that any chosen hadronic model must satisfy the nuclear bulk properties and all of them should provide macroscopic results within the observed NS mass and radius constraints.
}

\section{Results and analysis}\label{s3}

\begin{figure}
  \includegraphics[scale=0.5]{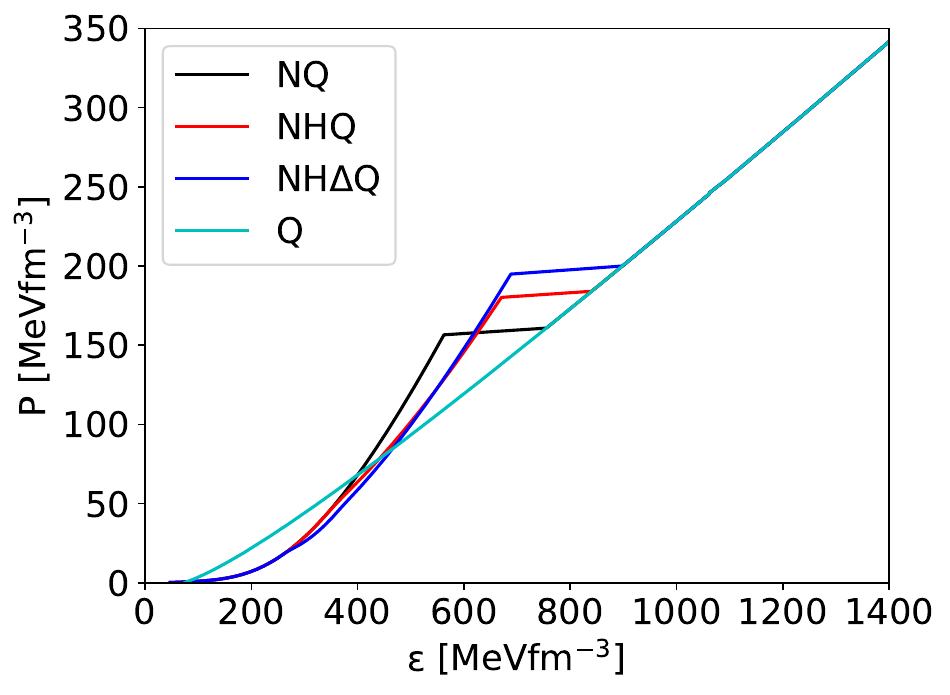}
\caption{We show the EoS for nucleons, nucleon plus hyperons, and nucleons plus hyperons plus $\Delta$-isobars admixture. The particles show sharp phase transitions at different energy densities represented by the discontinuity in the curves. Also, at lower densities, the quark matter shows similar behavior as the hadrons.}
    \label{EoS}
\end{figure}

In Fig.~\ref{EoS}, we present the EoS for the hybrid NS. The hadronic EoSs, composed of different particle contents, show phase transitions at different densities. The smaller hadronic critical chemical potential, $\mu_c$, is $1364\,\text{MeV}$ corresponding to the stiffer hadronic EoS composed of nucleons. Adding hyperons to the nucleon matter increases the $\mu_c$, as shown in Tab.~\ref{ma}, softening the EoS and delaying the phase transition. Again, including the $\Delta$-isobars further softens the EoS at low densities and increases $\mu_c$. Hence, the higher the $\mu_c$, the softer the hadronic EoS involved. Additionally, chemical potentials between 949 MeV and $\sim$1364 MeV, the quark matter shows characteristics similar to the hadronic matter phase. In the hadron-to-quark phase where the curves for $P$ and $\mu$ first intersect, as shown in Fig.~\ref{fig1}, the EoS shows a sharp discontinuity as shown in Fig.~\ref{EoS}. 

It has almost become standard these days to consider NSs with the entire spin-1/2 octet,
{since the hyperon puzzle can be circumvented in different ways.}
Lately, introducing the $\Delta$-isobars in addition to the baryon octet is actively being investigated at zero temperature \cite{Marquez:2022gmu, Li:2019tjx, Schurhoff:2010ph, Zhu:2016mtc}, and at a finite temperature and entropy \cite{ Sedrakian:2022kgj, Malfatti:2019tpg, Issifu:2023qyi}. Even though the new degrees of freedom are expected to soften the EoS and reduce the maximum NS mass, adjusting the baryon-meson coupling of the non-nucleonic components of the stellar matter with experimental and astrophysical data within the RMF approximation reduces this effect \cite{Weissenborn:2011kb, Lopes:2022vjx}. We observe that the hadron-quark phase transition does not resolve the so-called ``hyperon puzzle'' as can be seen from our results and also pointed out in \cite{Lopes:2020rqn}.

\begin{figure}
  \includegraphics[scale=0.5]{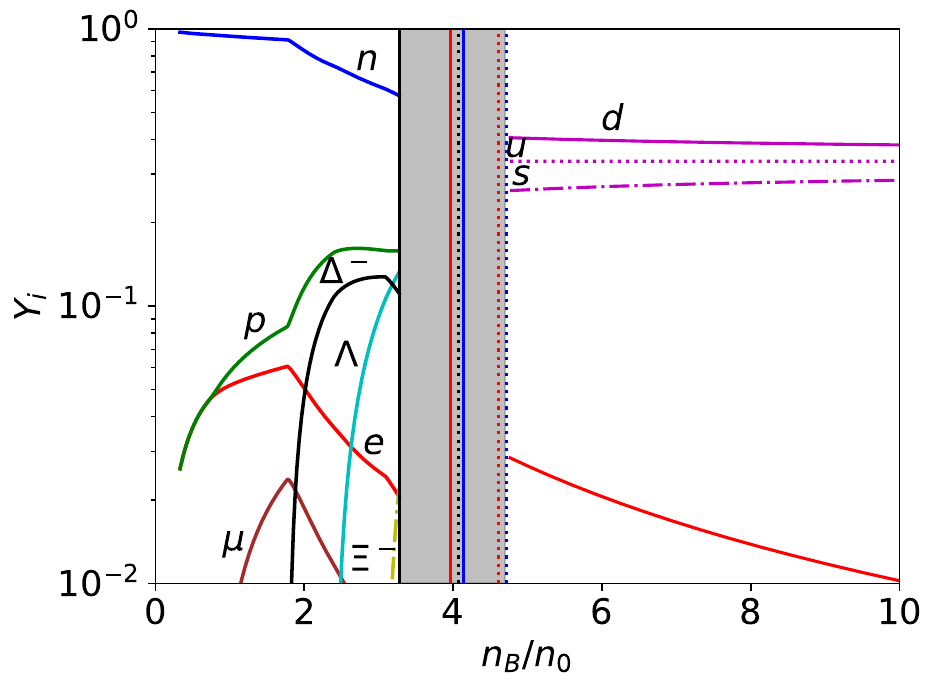}
\caption{The entire region shaded gray is where the hadron-quark phase transition for NQ, NHQ, and NH$\Delta$Q takes place. The black vertical line is where the N-quark phase transition starts, the red vertical line is where the NH-quark phase transition starts, and the blue vertical line in the gray area is where the NH$\Delta$-quark phase transition starts. While the corresponding dotted lines are the point where the phase transition is complete for each system of particles.}
    \label{pf}
\end{figure}

In Fig.~\ref{pf}, we show the particle composition, given by (\ref{yb}) applicable to both the quarks and the baryons of the stellar matter before, during, and after the hadron-quark phase transition. Before the phase transition, the star is composed of hadrons. As can be observed in Fig.~\ref{pf}, the star is mainly composed of non-strange baryons at densities below $n\sim 2.3~n_0$ before the strange baryons start showing up at about $2.5~n_0$; with the $\Lambda^0$ being the most dominant. The $\Delta$-isobars, on the other hand, appear at densities lower than $2~n_0$.
We have used black, red, and blue vertical lines to show where the hadron-quark phase transition starts (solid lines) and ends (dotted lines). Generally, we observe that adding new degrees of freedom to the stellar matter delays the hadron-quark phase transition. At densities of about $5~n_0$, the core of the star is mainly composed of quark matter, with $d$-quark being the most dominant. However, the $s$-quark rises steadily with density while the $d$-quark decreases slowly with density. 

{As we do not have a mixed phase, the particle fractions are calculated independently for each EoS. This way, we have a discontinuity in the electron chemical potential $\mu_e$ and in the energy density $\varepsilon$ at the point of transition, as can be seen in Fig.~\ref{EoS} and \ref{pf}, respectively. Fig.~\ref{pf} clearly shows the discontinuity in the electron fraction in the quark region (marked red after the phase transition) and the one in the hadron region (marked red before the phase transition). These discontinuities are the ones expected when a Maxwell construction is performed.}

\begin{figure}
  \includegraphics[scale=0.5]{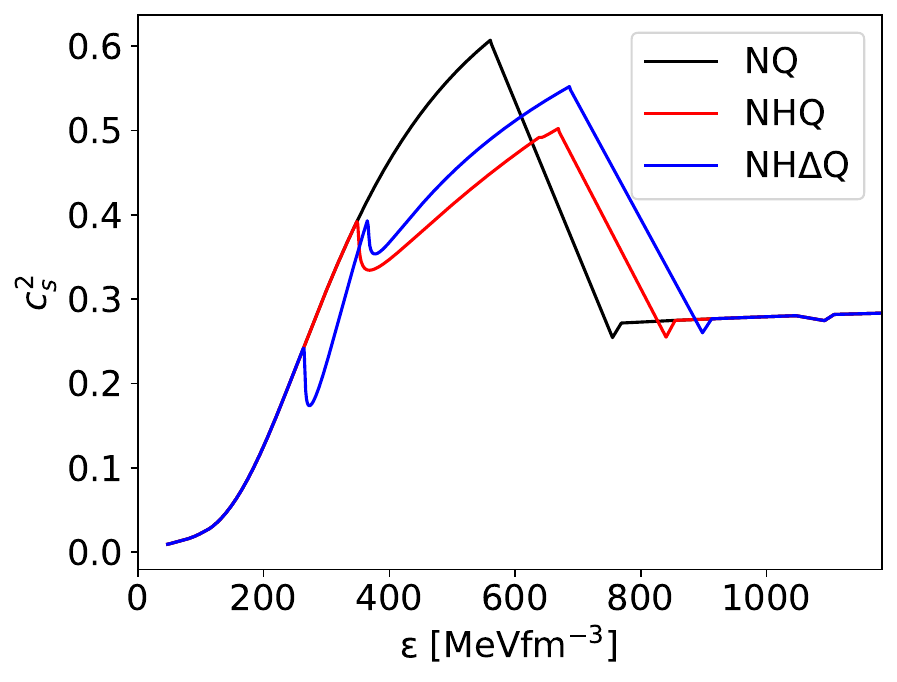}
\caption{ Square of speed of sound, $c_s^2$, as a function of energy density. The bumps in the curves at the lower densities represent the appearance of new particles. For instance, the curve for NQ shows no bump at the low density, NHQ shows two bumps for N and H and NH$\Delta$Q shows three bumps for each class of particles. At the higher density limits, there is a slight bump showing the appearance of the strange quark as well.}
    \label{fig2}
\end{figure}
\begin{figure}
  \includegraphics[scale=0.5]{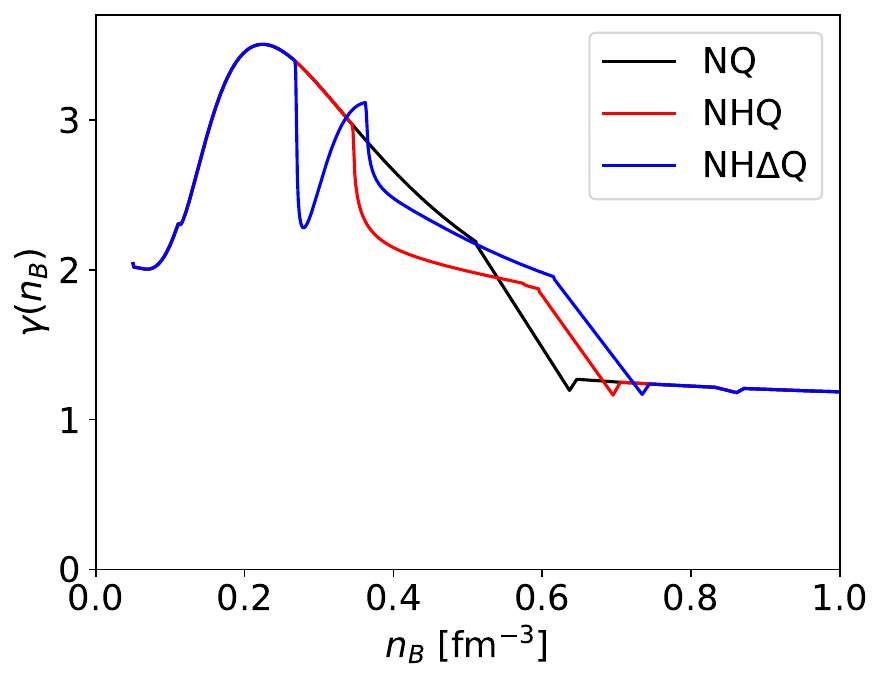}
\caption{Polytropic index as a function of the baryon density}
    \label{fig3}
\end{figure}

The perturbative and nonperturbative regions of the QCD theory with quark and hadron degrees of freedom exhibit different properties, respectively. Quark matter is approximately invariant under conformal symmetry, and hadron matter is not conformally invariant due to chiral symmetry breaking. These qualitative differences; can be observed by determining the values of some physical quantities, such as the speed of sound, polytropic and adiabatic indices, among others. 
Here, we will analyze the speed of sound $c^2_s\,=\,\partial P/\partial \varepsilon$ and the polytropic index $\gamma\,=\, \partial\ln P/\partial\ln\varepsilon$, 
in Figs.~\ref{fig2} and \ref{fig3}, respectively. The speed of sound is constant $c^2_s=1/3$ in the exactly conformal matter, and it approaches this value from below at high-density quark matter region \cite{Kurkela:2009gj}. The $c_s$ informs us about the star's internal dynamics and composition using the stiffness of the corresponding EoS. Hence, the appearance of new degrees of freedom leads to different behaviors of the $c_s$ in the stellar matter. 

The chiral effective field theory calculations show $c^2_s\,\ll\,1/3$, below the saturation density, whereas most hadronic matter at higher densities predicts $c^2_s\,\gtrsim\,0.5$ \cite{Bedaque:2014sqa, Annala:2019puf}. These qualitative predictions are in good agreement with the result presented in Fig.~\ref{fig2}. 
The $c_s$ grows monotonically with density in the hadron region. However, the onset of new degrees of freedom, such as hyperons and $\Delta$-isobars, leads to a sudden break in the monotonic behavior. A similar analysis was done in \cite{Lopes:2022vjx} using the adiabatic index.
The little bump at the high-energy region, where quark matter is found, shows the appearance of a strange quark. Looking at Figs.~\ref{fig2} and \ref{fig3} the appearance of $\Delta^-$ and $\Lambda^0$  at about $1.8~n_0$ and $2.5~n_0$ is immediately evident leading to the drop in $\gamma$ and $c_s$.

It has been argued that the characteristics of the $c_s$ are associated with the size of the quark core in the hybrid star. As shown in \cite{Annala:2019puf}, the $c_s$ in quark matter is related to the mass and radii of the investigated hybrid star. The authors showed that if $ c^2_s \,<\,1/3$, a massive NS is expected to have a massive quark core. From our model framework, we found that the most massive NS star is the one composed of only nucleons at low densities (see Tab.~\ref{ma}), in which case the quark core is responsible for 11\% of the mass of the star.

Moreover, the polytropic index, $\gamma$, attains a value of $\gamma = 1$ in the conformal matter region whereas, chiral effective field theory calculations and hadronic models predict $\gamma\approx 2.5$ around 
the saturation density \cite{Annala:2019puf}. In Fig.~\ref{fig3}, $\gamma$ starts rising in the hadronic matter region, with $\gamma\approx 2$ around the saturation density, until it peaks at $\gamma\approx 3.5$.
Then, $\gamma$ starts dropping to $\gamma\approx 2$ before it drops sharply, due to the hadron-to-quark phase transition at higher density regions, until it reaches $\gamma\approx 1$ in the quark matter phase, where the matter is expected to be conformal. Indeed the value of $\gamma$ coincides with the theoretical prediction for the quark matter region, and the maximum value obtained in the hadronic matter region is slightly higher than the chiral effective field theory prediction.

\begin{figure}
  \includegraphics[scale=0.5]{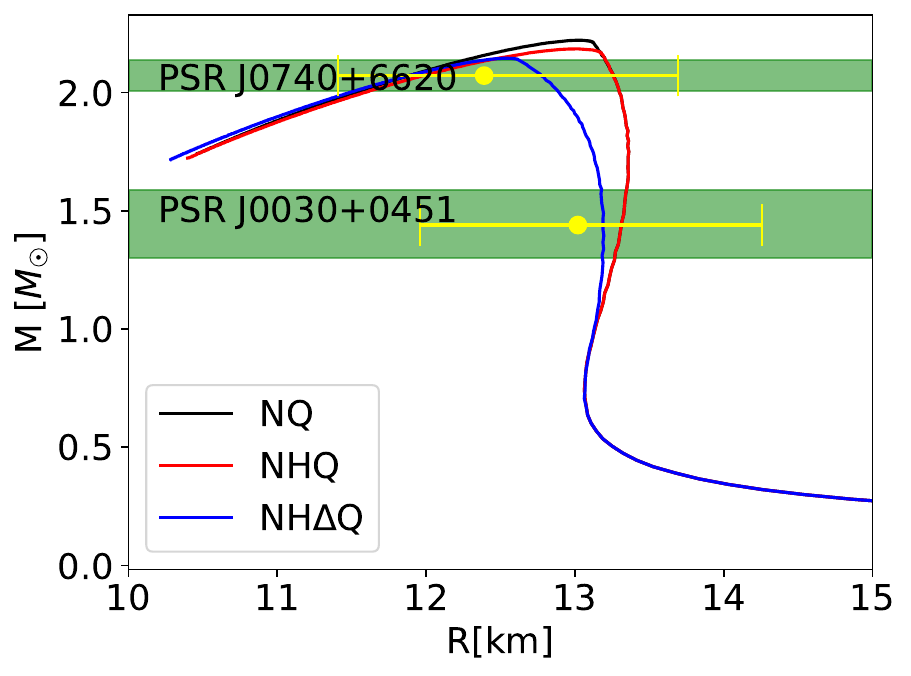}
\caption{In this diagram we show the  mass of the star as a function of its radius for each particle composition. { The green rectangles represent the masses of recent NICER NS with their error margins while the yellow points represent the radii and respective error bars.}}
    \label{mr}
\end{figure}

In Fig.~\ref{mr}, we use the EoSs generated from the study to calculate the mass and radii of the star using TOV equations \cite{PhysRev.55.374}. Also, we added BPS to the EoSs to simulate the NS crust \cite{Baym:1971pw}. It is important to mention that the mass-radius diagram is sensitive to the particle content of the star at the medium to the central part of the star. As expected, the addition of new degrees of freedom to the stellar matter softens the EoS and, as a consequence, reduces the maximum mass of the star \cite{Marquez:2022gmu, Raduta:2021xiz, Li:2019tjx}. The results presented in the figure above and Tab.~\ref{ma} are well within the $2\,\text{M}_\odot$ constraints imposed on NSs \cite{Antoniadis:2013pzd}. Recent data from NICER \cite{NANOGrav:2019jur, Fonseca:2021wxt} measure massive pulsar PSR J0740+6620 of mass $2.072^{+0.067}_{-0.066}\,\text{M}_\odot$ and radius $12.39^{+1.30}_{-0.98}$km at $68\%$ certainty \cite{Riley:2021pdl}. {The PSR J0030+0451 pulsar was also previously determined to have mass $1.44^{+0.15}_{-0.14}\,\text{M}_\odot$ and radius $13.02^{+1.24}_{-1.06}$km within the same confidence margin \cite{Miller:2019cac}.} Hence, there is a well-determined mass-radius window within which NSs can be described. Therefore, the model under discussion accommodates the description of PSR J0740+6620  and PSR J0030+0451 with hyperons, delta particles, and quark matter in its core.

Hypermassive stars with large quark cores were obtained in \cite{Lopes:2021jpm}, using the vector MIT Bag model and quantum hydrodynamic models, where the authors constructed hybrid stars with more than 80\% quark core. However, with the DDQM and DDME2 models, we found a less massive quark core of about 11\% of the mass of the star for a stellar matter composed of nucleons (most massive among NQ, NHQ, and NH$\Delta$Q). Comparing our results with \cite{Lopes:2021jpm}. We found that their softest quark matter EoS, corresponding to a Bag constant ${\rm B}^{1/4}=195{\rm MeV}$, shows a hadron-quark phase transition at $\mu_c = 1266{\rm MeV}$ and $P_c = 110{\rm MeV}$. Meanwhile, our stiffer hadronic matter EoS, composed of nucleons, has a hadron-quark phase transition at higher values of $\mu_c$ and $P_c$, as presented in Tab.~\ref{ma}. Thus, we can infer that higher relative values of $\mu_c$ and $P_c$ imply less massive quark cores in a hybrid NS. 

\begin{figure}
  \includegraphics[scale=0.5]{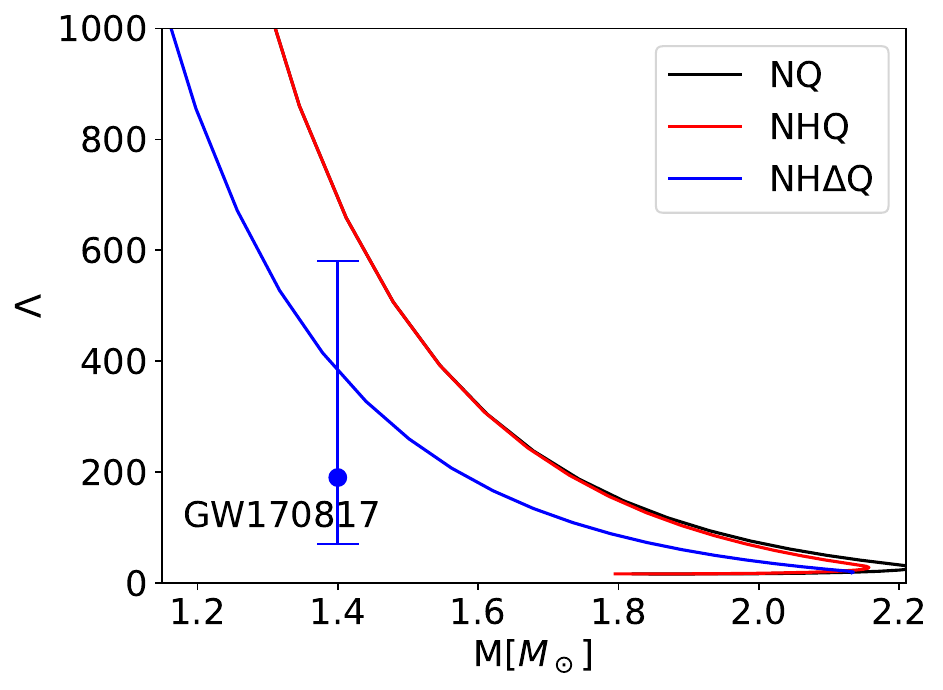}
\caption{The dimensionless tidal deformity ($\Lambda$) of a neutron star as a function of the star's mass. The point with the error bar represents a star mass $1.4{\rm M}_\odot$ that participates in the merger event.}
    \label{Td}
\end{figure}
{
The first detection of a binary NS merger, GW170817,
led to strong constraints on the NS masses and tidal deformability \cite{LIGOScientific:2017vwq,  LIGOScientific:2018hze}. 
Tidal deformability is a macroscopic astrophysical observable property of the NS that can be used to explore its interior. Similar to any external object with a defined structure, NSs can tidally deform when they come under the influence of an external tidal field. Through the coalescence signal of NS detected during the gravitational wave emission, the observed deformation can be quantified through the dimensionless tidal deformability parameter, $\Lambda$, which can be measured. In general, a larger value of $\Lambda$ signifies that the star is larger, less compact, and easily deforms. On the other hand, a smaller value of $\Lambda$ signifies a smaller star, highly compact and difficult to deform. Mathematically, the dimensionless tidal deformability is related to the gravitational Love number $k_2$, mass (M), and the radius (R) of the star as
\begin{equation}
    \Lambda = \dfrac{2}{3}k_2\dfrac{R^5}{M^5}.
\end{equation}
Thus the internal structure of the NS is imprinted on the tidal deformability constraint through $R$, $k_2$, and $M$ which provides complementary information on the interior dynamics of the star relative to its radius. In some cases, $\Lambda$ is expressed in terms of the compactness of the star $C=M/R$.

In Fig.~\ref{Td} we show the results of the tidal deformability as a function of the star's mass calculated from the EoS for hybrid stars.{ One can see that only
the result with $\Delta$ baryons satisfy the constraint imposed by GW170817, i.e.,  $\Lambda = 190^{390}_{-120}$ \cite{LIGOScientific:2017vwq} whose signal range has been shown with an error bar in the figure.}
}

\section{Conclusion}\label{s4}
We studied the structure and composition of hybrid NSs constituted of nucleons, hyperons, and $\Delta$-isobars admixed hypernuclear matter with a quark core. We constructed our EoSs for the study using the Maxwell method with a sharp hadron-quark phase transition. We determined the properties of the hybrid star, such as the EoS, $c_s$, $\gamma$, $M_\text{max}$, $R_{\text{max}}$ and $Y_i$, and compared the results with both theoretical and observational data. Some of the results are listed below:
\begin{itemize}
    \item We determined a hybrid NS with a maximum mass within the $2\,\text{M}_\odot$ threshold, composed of the baryon octet, $\Delta$-isobars, and a quark core.\\
    \item We observed that hybrid NSs are more likely to be composed of non-strange baryons at low baryon densities, while baryons with strangeness are found towards the center of the star.\\
    \item We established a relationship between the $\mu_c$, hadron-quark phase transition, softening of the EoS, and the composition of particles in the star. Higher values of $\mu_c$ imply softer EoS, delay in the phase transition, and higher particle degrees of freedom in the star (i.e., either nucleon plus hyperons or nucleon plus hyperons plus $\Delta$-isobars admixture).\\
    \item Also, the value of $P_c$ informs us about the phase transition, structure, and composition of the star. As can be seen in Tab.~\ref{ma}, higher $P_c$ implies softer EoS, low $M_\text{max}$, small $R_{\text{max}}$, higher particle degrees of freedom and a delayed phase transition, {as extensively discussed in the literature \cite{Lopes:2021jpm}.} \\
    \item The $c_s$ calculated from the model framework shows $c_s^2\,\approx\,0.6$ in the hadron region, and $c_s^2\,\approx\,0.3$ in the quark matter region; well within the $c_s^2\,\lesssim\, 1/3$ threshold for conformal matter. These results are in good agreement with the theoretical predictions \cite{Bedaque:2014sqa, Annala:2019puf}
    \\
    \item  The value of $\gamma$ determined in the model framework also shows good agreement with the theoretical predictions. In the quark matter region, we determined $\gamma\,\approx\,1$, and in the hadron region, we obtained a value slightly higher, $\gamma\,\approx\,3.5$, than the theoretical values predicted in the literature \cite{Annala:2019puf}.
\\
    \item {{We calculated the dimensionless tidal deformability and showed that only the results with $\Delta$ baryons satisfy the constraint imposed by GW170817 for all hybrid stars.}}
\end{itemize}
Our findings are comparable to previous works on hybrid NSs using different approaches. The characteristics of the $c_s$, $\gamma$, and the theoretical evidence of a quark core in dense NSs matter are extensively discussed in \cite{Annala:2019puf, Bedaque:2014sqa}. The approach adopted here for the determination of the hadron-quark phase transitions and the construction of the hybrid EoSs is similar to what can be seen in \cite{Lopes:2021jpm, Contrera:2022tqh}. { However, we have used density dependent models to describe both the hadronic and quark matter models.} {It is worth noting that current data available on NS observation measures only the maximum mass and radii and does not help us to discriminate between the particle content of the star. In this case, it would be impossible to define the inner star constituents, at least for now. However, with current theoretical advancements, the stars can be distinguished by determining parameters such as the $c_s^2$, $\gamma$, $M$, $R$, and $\Lambda$ among others. In the current work, the mass-radius diagrams are all very similar but the $c_s^2$ and $\gamma$ show clear differences among the stars in terms of their particle contents.}

{In the present work, the uncertainties in the hadronic phase were explored by analyzing three possible compositions for the nuclear matter to see which one would produce the best results. The DDME2 parameter set is the one that satisfies nuclear bulk properties and the hyperon-meson and $\Delta$-meson coupling constants were calculated by using SU(3) symmetry arguments, as proposed in \cite{Lopes:2022vjx} and already tested in \cite{Issifu:2023qyi}.
In the case of the quark phase, we tried to manually select the values of the parameters C and D that besides being slightly outside the stability window would also maximize the size of the quark core. Currently, we are working on a Bayesian approach to try to optimize the parameters of the DDQM model in view of the current astrophysical constraints, but this is still a work in progress.}

\section*{Acknowledgements}

This work is a part of the project INCT-FNA Proc. No. 464898/2014-5. D.P.M. was partially supported by Conselho Nacional de Desenvolvimento Científico e Tecnológico (CNPq/Brazil) under grant 303490-2021-7 and A.I. under grant 168546/2021-3. F.M.S. would like to thank FAPESC/CNPq for financial support under grant 150721/2023-4.
{We thank Dr. Luiz Laercio Lopes for fruitful discussions.}


\section*{Data Availability}
The datasets generated and/or analyzed during the current study are available from the corresponding author upon reasonable request.

\bibliographystyle{mnras}
\bibliography{references}

\label{lastpage}
\end{document}